# Représentation de données et métadonnées dans une bibliothèque virtuelle pour une adéquation avec l'usager et les outils de glanage ou moissonnage scientifique

G. Kembellec, *Laboratoire PARAGRAPHE, Université Paris 8*

*Abstract*— The vehicles for official knowledge, as well as university libraries, suffer from an increasingly visible lack of interest. This is due to the advent of fully digital practices. By studying the psychological and cognitive models in information retrieval initiated in the 1980s, it is possible to use these theories and apply them practically to the Information Retrieval System, taking into account the requirements of virtual libraries. New metadata standards along with modern tools that help managing references should help automating the process of scientific research. We offer a practical implementation of the given theories to test them when they are applied to the information retrieval in computer sciences. This case under study will highlight good practices in gleaning and harvesting scientific literature.

## I. Introduction

Les vecteurs officiels de la connaissance en milieu académique, comme les bibliothèques universitaires et leur catalogue, souffrent depuis l'avènement du tout numérique d'un désamour de plus en plus visible. Les cœurs de cible de ces organismes sont les étudiants, principalement en deuxième et troisième cycles, les enseignants et les chercheurs. Dans un même temps, les bibliothèques académiques s'équipent de technologies de pointe. Cependant, le résultat ne semble pas forcément en adéquation avec les besoins des utilisateurs.

Les centres documentaires, notamment universitaires, se sont adaptés à cette constatation en informatisant leurs catalogues et les interconnectant. Notons l'effort français de l'Agence bibliographique pour l'enseignement supérieur (ABES), relayée par les services documentaires des universités, pour numériser et cataloguer les thèses dans le projet titanesque Thèses En Ligne (TEL) et le dispositif STAR. Malheureusement, cette mise à disposition n'induit pas une consultation systématique des principaux intéressés. Selon Markey [1], les catalogues des bibliothèques sont tombés en disgrâce. Une étude de Rosa menée en 2006 pour le compte de l'*Online Computer Library Center* a démontré que rares sont les étudiants, enseignants et chercheurs commençant leurs recherches par le catalogue de leur bibliothèque[2]. Ils sont 89 % à lui préférer un moteur de recherche commercial, au premier rang desquels Google. Selon l'étude de 2005 de Swan & Brown, 72 % des universitaires utilisent le moteur Google. Lorcan Dempsey, directeur de la recherche à l'*OCLC*, a qualifié notre époque d'ère « Amazoogle [3] », faisant allusion au moteur de recherche Google et à la librairie numérique Amazone. Selon lui, il faut aller chercher le public là où il est, c'est à dire sur Internet. Ce constat de modification de pratiques des usagers de l'information (fut-elle scientifique) induit un changement dans les usages professionnels de la documentation, qu'ils soient documentalistes ou éditeurs de logiciels. Selon de Kaenel & Iriarte[4], « les dernières évolutions du web, avec l'entrée en jeu enfin de XML, des nouveaux usages et nouveaux outils, ainsi que le déplacement du centre de gravité qui s'est fortement rapproché des utilisateurs, ouvrent de nouvelles voies et de nouveaux champs d'application pour les catalogues en ligne ».

En étudiant les modèles cognitifs et psychologiques en recherche d'informations initiés depuis les années 1980, il est possible de dégager une méthodologie et de l'adapter à un *Système de Recherche d'Informations* (SRI) en adéquation avec les impératifs des bibliothèques virtuelles. Les nouvelles normes de métadonnées, associées aux outils modernes d'aide à la gestion de bibliographie doivent permettre d'automatiser le processus de recherche scientifique. Cette automatisation du cycle de recherche va de l'établissement d'un périmètre de recherche, jusqu'à la production de documents bibliographiques normés dans l'optique de la production scientifique

Dans cet article, nous allons dans une première partie faire un panorama des méthodes de recherche d'informations, en mettant l'accent sur les aspects cognitifs et psychologiques. Nous en déduirons un processus théorique complet de *Recherche d'Informations* (RI) que nous adapterons à un *Système de Recherche d'Informations*. Cette démarche nous offrira la possibilité de mettre en adéquation les bibliothèques numériques avec ses usagers devenus plus exigeants à cause de l'usage des outils offerts par les géants de la recherche d'informations.

Dans une dernière partie, nous proposerons une mise en pratique des théories dégagées pour en tester l'usage dans le cadre de la recherche d'informations en informatique.

II. LES MODELES DE RECHERCHE D'INFORMATIONS

*A. Gutherie, le modèle initial de la RI moderne*

Gutherie a le premier établi un modèle de l'activité cognitive lors de la recherche d'informations. Bien qu'ayant été révisée depuis, cette visualisation du processus de recherche a servi de base pour beaucoup de chercheurs. Cinq étapes ont été définies de la formalisation d'un but jusqu'au traitement de l'information. Tout d'abord, la personne qui commence une activité de recherche se fixe une mission et ce faisant il élabore une représentation de l'objectif à atteindre. Puis il convient de chercher la source d'information qui est censée apporter les réponses attendues. Ensuite, il faut synthétiser la documentation pour en extraire l'information. Enfin, l'information, si elle est pertinente, est intégrée, c'est-à-dire qu'elle est reliée aux autres savoirs que possède le sujet.

*B. « Information seeking process » de Marchonini*

Dans le troisième chapitre de son livre « *Information Seeking in Electronic Environments* », Marchionini propose une autre vision du processus de recherche d'informations étendant celui de Gutherie. Son approche méthodologique de recherche d'informations se divise en 8 points que nous détaillerons, car il s'agit d'une des méthodes tirées de Gutherie les plus abouties et reconnues.

1. Reconnaître et accepter un besoin d'information, simplement la prise de conscience d'une lacune.
2. Définir et comprendre la problématique de recherche. La définition de la problématique est une étape critique dans la recherche d'informations. Ce sous processus reste actif tant que la recherche avance. Marchionini insiste sur le fait d'avoir à redéfinir la problématique au fur et à mesure de l'avancement (et donc de la compréhension) du sujet d'information progresse.
3. Le choix d'un système de recherche dépend de l'expérience du demandeur d'informations. Le type des réponses attendues lors de la définition de la problématique de recherche sera un facteur de choix. La qualité des réponses n'est pas la même sur Wikipédia et sur un portail scientifique. Plus l'utilisateur est affûté dans son domaine plus son portail de prédilection lui sera familier et moins il sera enclin à utiliser un service générique.
4. La formulation de requêtes implique la compréhension de la syntaxe de recherche. La formulation la première requête sert de point d'entrée. Cette étape est suivie par une ou des reformulations de la requête. Cela implique une bonne connaissance de la sémantique du domaine et de son champ lexical qu'il s'agisse d'un portail de recherche ou d'un moteur. De plus, l'utilisateur doit parfois maîtriser le langage booléen de requête utilisé par l'interface pour affiner sa requête.
5. Cette étape est décrite comme la simple aptitude à utiliser un hypertexte, ce qui n'est plus vraiment d'actualité. En effet, toute personne connaît ce type d'usage.
6. La réponse fournie par le système est un résultat intermédiaire qu'il faut examiner avec la plus grande attention. Selon la quantité et la qualité des réponses obtenues par l'usager, ce dernier peut évaluer ses progrès dans le processus global, mais également dans l'activité de RI. Il est probable qu'il faille à cette étape relancer une requête (étape 4) ou redéfinir le problème (étape 2).
7. Pour extraire des informations, un utilisateur de SRI use de compétences comme la lecture rapide, l'analyse et la classification. Le chercheur doit entre capable de citer de manière clairement identifiable (synthèse de texte) et de référencer la citation dans une bibliographie.
8. Une recherche d'informations aboutit rarement en une seule requête. Le plus souvent, le premier résultat sert de base pour la formulation et l'évaluation de requêtes supplémentaires. Décider d'arrêter la recherche exige une évaluation positive du processus complet. Il arrive régulièrement à cette étape de devoir réexaminer les résultats (étape 6), peut-être de relancer une requête (étape 4), voire dans certains cas de redéfinir sa problématique de recherche (étape 2).

*C. Le modèle ISP de Kulthau*

Carole Kulthau met en évidence les étapes du processus de recherche d'informations tout en y associant des sentiments, pensées, actions, et tâches. Dans son livre Seeking Meaning, Kuhlthau [6] énonce le principe suivant :

« Le processus de recherche d'informations est enclenché par un état d'incertitude dû à un manque de compréhensions, à un sens inexpliqué, à une structure incomplète. Il s'agit d'un état de nature cognitive qui provoque généralement des symptômes affectifs comme l'anxiété et le manque de confiance[1]».

Les principes initiaux de cette théorie avaient été énoncés dans un article de 1991 dans lequel Kuhlthau énonçait un probable parallèle entre l'évolution de la qualité la recherche d'informations et l'expérience du sujet (le jeune chercheur) et son état émotionnel [7]. Son approche est pédagogique, son contexte de recherche se situe donc dans le cadre de la formation de chercheurs apprenants. Cependant, lorsqu'un chercheur, même expérimenté, s'aventure sur un domaine qu'il ne maîtrise pas pour une recherche transverse, nous supposons que le processus doit être identique.

1. Au commencement, une personne prend connaissance d'un manque, que ce soit en matière de connaissance ou de compréhension. La personne doit faire face à de sentiments d'incertitude et d'appréhension. À ce stade, la tâche est simplement de reconnaître un besoin d'information. De manière diffuse, le chercheur voit naître des réflexions sur la manière d'envisager le problème, de comprendre la tâche. Pour faire face à cette problématique, il doit essayer d'en faire le lien avec ses expériences et connaissances antérieures.
2. Au cours de la « sélection », la tâche est d'identifier et de sélectionner le thème général sur lequel effectuer sa recherche et de l'approche à adopter. Le sentiment d'incertitude cède alors fréquemment le pas à l'optimisme. Les réflexions sont alors centrées sur le poids des termes clés à chercher par critères d'intérêt personnel et d'informations disponibles. Certains peuvent effectuer un balayage visuel sur une recherche préliminaire pour un aperçu de mots clés alternatifs. Lorsque, pour une raison quelconque, cette sélection est retardée ou reportée, les sentiments d'anxiété sont susceptibles de s'intensifier jusqu'à ce que le choix des termes soit arrêté.
3. L'étape d'exploration se caractérise par des sentiments de confusion. L'incertitude et le doute augmentent souvent chez le chercheur lors de cette période. L'activité consiste à étudier le thème général en vue d'accroître la compréhension du domaine. Ces réflexions permettent au chercheur d'être suffisamment éclairé sur le sujet pour se constituer un point de vue. À ce stade, l'incapacité de l'utilisateur (s'il est débutant) à formuler précisément une requête rend la recherche malhabile. Cette partie comporte une phase de la localisation d'informations sur le sujet général. La lecture d'informations générales permet de se renseigner sur la thématique, l'affinage amène de nouvelles informations qui viennent s'agréger à ce qui est déjà connu. Il semble plus pertinent pour Kuhlthau de laisser l'utilisateur recouper ses informations et les synthétiser pour laisser émerger des schémas de pensées. Cependant, les informations rencontrées correspondent rarement à un schéma unique. Les informations provenant de différentes sources semblent souvent contradictoires, voire incompatibles du fait de l'évolution des recherches et des courants de pensée. Les utilisateurs peuvent trouver la situation tout à fait décourageante et se sentir angoissés et frustrés. Certains chercheurs sont peut-être enclins à abandonner la recherche à ce stade.
4. La formulation est le pivot du processus de recherche selon Kuhltau, lorsque la confiance en soi et le système augmente. Grâce à l'ensemble des documents parcourus, une idée directrice du domaine de recherche prend forme. Cette réflexion implique d'identifier et de sélectionner des concepts clés qui mettent en perspective les termes prédominants autour desquels la collecte d'information va s'organiser. La formalisation des sujets va permettre au chercheur de s'approprier les concepts par une reformulation personnalisée.
5. La « collection » intervient au moment où l'utilisateur maîtrise les fonctions du système d'information. À ce stade, il est à même de recueillir des informations liées au thème ciblé par une recherche exhaustive de toutes les ressources disponibles. Le sentiment de confiance continue de croître avec l'approfondissement de la connaissance du sujet.
6. L'étape de « présentation » est la dernière du processus. Il s'agit de mettre au propre les notes et les brouillons et se préparer à exploiter les résultats. Les doublons et le bruit sont éliminés. Cette dernière étape s'accompagne de sentiments de soulagement et de satisfaction si la recherche s'est bien déroulée (ou de déception si elle n'a pas abouti).

---
[1] Traduction de Paulette Bernhard, professeure à l'École de bibliothéconomie et des sciences de l'information (EBSI), Université de Montréal.

Si l'approche de Kuhltau est très rigide, voire scolaire, l'analyse des sentiments associés aux étapes est d'un intérêt incontestable. En effet, ce retour d'observation offre un aperçu séquentiel des émotions et de la psychologie lors d'une recherche d'informations. Il y a une leçon à en tirer lors de la modélisation d'un SRI dans l'optique d'augmenter les chances de le rendre en adéquation avec son public.

### D. Le modèle « berrypicking » de Bates et le phénomène de sérendipité

Marcia J. Bates a publié de nombreux ouvrages dans les domaines de la stratégie de recherche du système d'information, la conception centrée sur l'utilisateur des systèmes de recherche de l'information. Très intuitif, le modèle de Bates est appelé « berrypicking » [8], car il compare la recherche d'informations à une cueillette de baies éparpillées dans les « buissons de connaissance ». Dans un processus de recherche documentaire, le chercheur consulte un premier document qui lui offrira de nouvelles pistes de recherche d'informations, que ce soit des idées ou d'autres références. Les idées soutenues dans le premier document mèneront à d'autres axes de recherches ou points de vues sur lesquels il faudra se documenter. Cette méthode introduit le phénomène de sérendipité. La sérendipité est le don ou la faculté de trouver quelque chose d'imprévu et d'utile en cherchant autre chose, ou encore, l'art de trouver ce qu'on ne cherche pas.

Le chercheur se constituera rapidement une arborescence documentaire, qui pourra même devenir un maillage, car les liens bibliographiques finissent par se recouper. Ce modèle permet de constituer rapidement une bibliographie autour d'un concept. Cependant, l'abondance d'informations et de documents peut également noyer le chercheur dans un flot d'idées contradictoires et le rendre indécis sur sa recherche au lieu de le conforter dans l'exploitation de son idée originale. Cette méthode est cependant excellente pour développer la faculté de sérendipité, indispensable pour étoffer sa RI. Cependant dans leur livre Van Andel & Bourcier [9], font bien le distinguo entre chance et pugnacité. La sérendipité n'est donc pas le hasard. Il s'agit d'un usage continu de la sagacité et de la perspicacité du chercheur (... d'information). Il s'agit plus de recouper des idées et les bibliographies pour extraire la « pépite » de connaissance. Il peut arriver en cherchant à comparer deux courants de pensée à en faire émerger un troisième moins connu. Il faut cependant maîtriser la chaîne de recherche pour prétendre faire émerger quoi que ce soit d'un processus documentaire. Nous pouvons dire que dans le cadre du modèle de Bates la recherche documentaire avance par « saut » d'un document vers un autre. Le lien motivant l'analyse du document suivant n'est pas une évolution séquentielle classique comme un classement dans

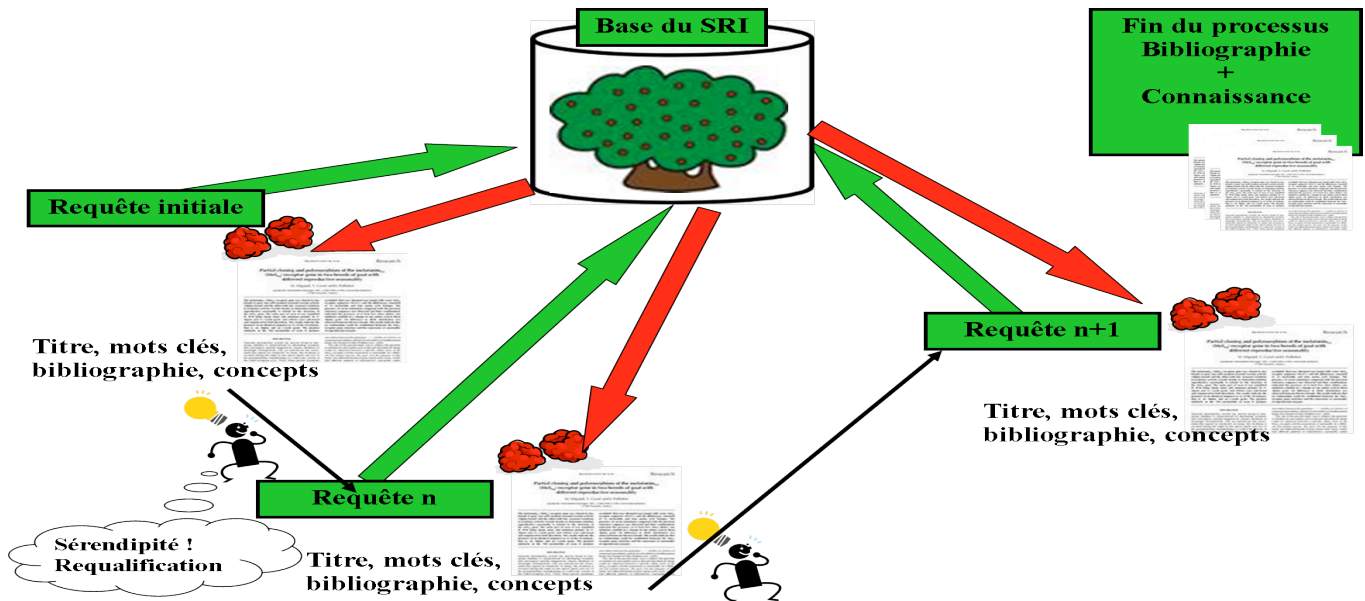
Figure 1. Modèle *"berrypicking"* de Bates.

un moteur, mais la mise en exergue d'une référence bibliographique rencontrée à plusieurs reprises, un hyperlien, une métadonnée (mot clé) ou une idée novatrice.

### E. Synthèse et critique des méthodologies

Au vu des quelques éléments que nous venons d'examiner, il est possible de classifier les modèles de recherche d'informations en 3 catégories.
– Recherche organisée, planifiée, de manière séquentielle ;
– Recherche par tâches ;
– Recherche itérative par sauts.

Le modèle de base en recherche d'informations est souvent celui proposé par Guthrie. D'autres chercheurs, plutôt que de vouloir l'approfondir ou le corriger comme Marchionini ont proposé une approche totalement différente. Bates propose l'abandon du caractère séquentiel de la recherche au profit d'un processus plus itératif, basé sur l'intuition. D'autres, comme Kuhltau, abordent une piste transverse. Selon la thèse de Kuhltau, d'une manière générale le sujet lors d'une RI passe de l'incertitude à la sérénité, à la condition, bien sûr, que le travail progresse correctement et que les informations trouvées le satisfassent. Cette optique met en avant l'aspect émotionnel de la recherche en balisant la méthode de recherche pour une recherche « sereine ». Entre les étapes de « formulation » et de « collection », Kuhlthau insiste sur la dépendance entre maîtrise du système d'interrogation et du domaine de connaissance.

Pour nuancer ces propos il faut distinguer que l'expertise en recherche d'informations se mesure distinctement sur ces deux axes. En effet, si la connaissance du domaine de recherche est indispensable à une recherche efficace, la bonne maîtrise des outils de recherche est également d'une importance capitale. Dans une étude de 2004 portant sur une cinquantaine d'étudiants, les chercheurs Ihadjadene & Martins [10] montrent l'équivalence entre expertise du domaine et familiarité dans l'usage du Web. Ainsi, dans cette étude, la moitié des sujets sont experts en psychologie et l'autre en d'autres matières. Dans chaque groupe, la moitié est très à l'aise avec les techniques de recherche sur Internet, l'autre peu à l'aise. Les étudiants qui ont la double compétence de la connaissance du domaine et de la maîtrise d'Internet présentent de meilleures performances. Ceux qui ne présentent qu'une des deux, peu importe laquelle, ont des résultats identiques.

Dans l'optique de réaliser un système optimisé de recherche documentaire, nous reprendrons les conclusions de l'étude de Rouse & Rouse [11] relatives à la littérature de l'information sur les comportements de recherche : « *Because information needs change in time and depend on the particular information seeker, systems should be sufficiently flexible to allow the user to adapt the information seeking process to his own current needs. Examples of such flexibility include the design of interactive dialogues and aiding techniques that do not reflect rigid assumptions about the user's goals and style* ». Nous retiendrons le caractère évolutif des besoins d'un individu en matière de recherche d'informations. C'est pourquoi les systèmes doivent être suffisamment souples pour permettre à l'utilisateur d'adapter le processus de recherche de l'information à ses propres besoins. Nous retiendrons donc que ce SRI se doit d'être un outil d'apprentissage du domaine tout en restant abordable au sens de Kuhlthau. En effet, en tenant compte des sentiments liés aux étapes de la recherche, nous éviterons de décourager le demandeur d'information.

### III. PROTOCOLES ET OUTILS DE RECHERCHE D'INFORMATIONS

Selon Bates [12], beaucoup d'utilisateurs, sinon la plupart, des systèmes d'information automatisés veulent profiter de la vitesse et la puissance de recherche automatique, tout en contrôlant et en dirigeant les étapes de la recherche eux-mêmes. Ils ne veulent pas que le système prenne en charge la recherche entièrement à leur place. Le juste milieu est d'allier la puissance de calcul à l'intuition et l'intelligence humaine, si possible collective. L'équilibre s'obtiendra en facilitant le travail du chercheur, c'est à dire en assistant la requête, en fournissant un résultat pertinent, avec un minimum de bruit et de silence. Ainsi, par abduction, faculté de psychologie cognitive dure à reproduire par un automate, l'utilisateur sera à même de faire émerger la documentation nécessaire à son travail de recherche. Ce postulat de départ énoncé, le résultat n'est exploitable que s'il est compatible avec les outils usuels du chercheur. Cette deuxième partie s'articulera dans un premier temps autour de la présentation des protocoles et formats couramment usités sur internet en RI. Ensuite, nous proposerons une synthèse critique et objective des outils les plus affûtés pour la RI scientifique.

### A. Protocoles et format d'échanges de données bibliographiques de la Websphère

#### 1. Format orienté moissonnage

Le terme de moisson en science de l'information est issu du terme anglais de science de la documentation « *harvesting* », qui signifie récolte ou moisson. Cette métaphore est particulièrement adaptée, car un « moissonnage » d'articles scientifiques est par défaut une recherche au sens large, à faible granularité comme une « récolte » au sens agricole. Le traitement de l'information, qu'il soit manuel ou automatisé, est réalisé a posteriori, comme l'agriculteur qui va « séparer le bon grain de l'ivraie[2] » après la récolte.

L'*Open Archives Initiative Protocol for Metadata Harvesting (OAI-PMH)*, ou protocole de collecte de métadonnées de l'Initiative des Archives Ouvertes propose un standard relativement simple pour l'échange de métadonnées. Créé par l'Initiative des Archives Ouvertes en 1999[3], il avait pour objectif premier de « faciliter la description et la diffusion des métadonnées d'articles scientifiques disponibles en accès ouvert sur Internet [13] ».

Pour être plus précis, un moissonnage va, le plus souvent offrir l'accès à l'intégralité des fiches documentaires d'un ou plusieurs entrepôts de données spécifiques. Il arrive que certains entrepôts de données puissent effectuer un prétraitement (cf. Fig. 2). Pour ce faire, les entrepôts offrent une URL OAI qui tient compte du sujet de la moisson en analysant la requête intégrant quelques mots clés et en les projetant sur les métadonnées contenues dans les notices.

#### 2. Formats orientés glanage

Pour poursuivre l'analogie avec le champ lexical agricole, ce qui est décrit ici comme du glanage contextuel est une méthode alternative de recherche de documentation scientifique. Le glanage contextuel est la possibilité de repérer de la documentation scientifique grâce aux métadonnées des contenus sur Internet. Il peut s'agir d'une cueillette manuelle, avec une intégration manuelle des métadonnées dans un format visible et compatible, comme le BibTeX, le RIS ou s'agit d'une urbanisation du système d'information avec les solutions logicielles grâce aux formats compatibles

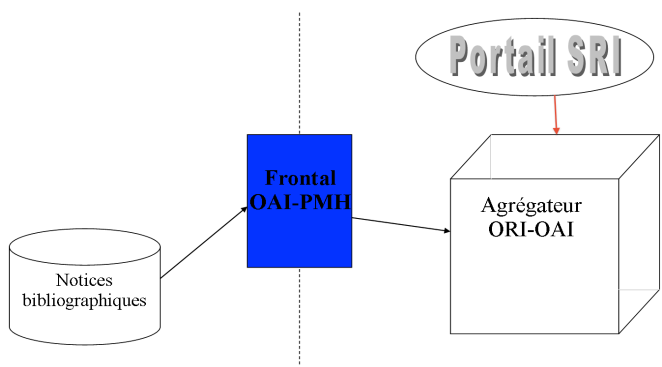

**Figure 2. Schéma de moissonnage ORI-OAI**

(OpenURL / COinS, Micro-format / RDFa, Gridll, métadonnées XHTML Dublin Core…).

### B. Logiciels à utiliser en collaboration avec un SRI pour une recherche optimisée

Un logiciel de gestion bibliographique est destiné à établir, trier et utiliser des listes de citations relatives à des revues, des articles, des sites web et des livres. Ce type de logiciel est principalement utilisé par les étudiants et enseignants et chercheurs de l'enseignement supérieur. Son usage est principalement adapté à la publication scientifique. Ces logiciels intègrent généralement une base de données. Cette dernière peut s'alimenter de différentes façons. Le cadre classique est l'interrogation des serveurs de connaissances scientifiques par recherche massive (moissonnage) ou en détectant les métadonnées dans les documents (glanage). Un autre usage est l'intégration manuelle de notices. Avec la base interne de ces logiciels, il est parfois possible d'effectuer une sélection de documents par recherche à facette. Les programmes de ce type se distinguent souvent par leur capacité à importer et exporter des formats de métadonnées informatiques différents reconnus comme RIS ou BibTeX. Pour tester ces logiciels, nous avons créé une page glanable et moissonnable contenant 3 entrées bibliographiques exposées visuellement et en métadonnées aux formats OpenURL / COinS (norme Z39 88) et BibTeX[4].

#### 1. Mendeley Desktop

Mendeley Desktop permet d'importer directement les métadonnées depuis les documents, notamment les fichiers PDF. Les autres formats d'import sont le Ovid, le RIS, le BibTEX, le EndNote XML, le TXT (texte brut), mais aussi, et c'est un point très appréciable, la base SQLlite du plug-in Zotero de Firefox. Ce logiciel offre une gestion des fichiers PDF performante capable de lire les métadonnées des documents afin d'auto générer les notices bibliographiques. Mendeley est compatible avec les logiciels MS Word et OpenOffice pour l'intégration de bibliographies. Un plug-in bookmarklet intégrable au navigateur (via les marques pages) permet de détecter les notices au format COinS et de les intégrer.

#### 2. JabRef

JabRef est une interface graphique java de recherche d'informations scientifique et de gestion de bibliographie. Ce logiciel permet de maintenir exclusivement des bibliographies au format BibTeX. Les principaux formats d'export normés supportés sont le RIS, le MODS, le RDF. JabRef est directement interconnecté avec des bases de connaissances scientifiques comme CiteSeer, IEEEXplore, JStor, ArXiv.org et autres portails ACM. Cela autorise des moissonnages massifs. Outre ces quelques formats d'export classiques, JabRef offre la possibilité de se connecter à une base de données et d'exporter sa bibliographie sous MySQL. Cette fonctionnalité est particulièrement attractive après un moissonnage ou un glanage intensif d'information avec export en format normé. Le processus est donc presque automatisé entre navigation avec le plug-in Firefox Zotero et la génération d'une base de connaissances de tous les articles intéressants sur un sujet. Il suffit pour cela de configurer une connexion à un serveur de base de données MySQL.

---

[2] Dans cette parabole de l'évangile de Mathieu, un ennemi a semé de l'ivraie dans un champ de blé, le maître dit à ses serviteurs de ne surtout pas chercher à l'enlever tant que la moisson n'est pas prête, sinon ils risqueraient d'arracher également le bon grain. Il leur demande donc d'attendre le bon moment, de ramasser alors l'ivraie pour la faire brûler puis de moissonner le blé pour le ranger dans le grenier.

[3] http://www.openarchives.org

[4] Disponible en ligne : http://ontologynavigator.paris-sorbonne.fr/biblio/

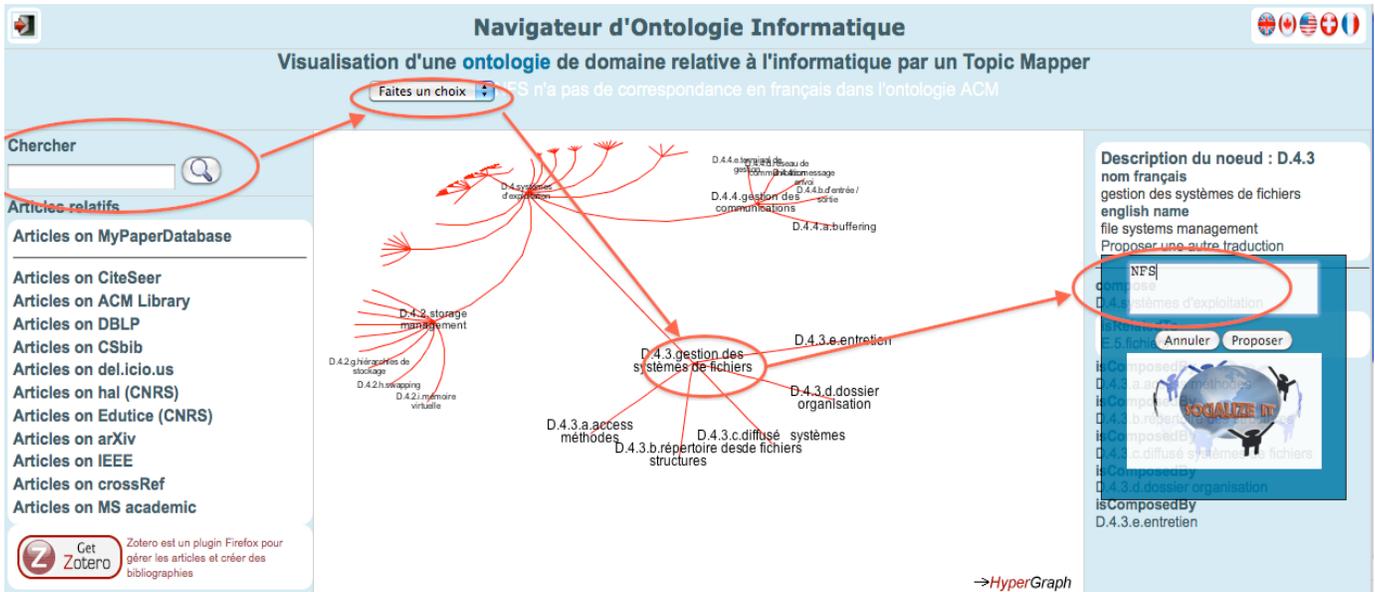

**Figure 2. SRI adaptable à l'utilisateur et au groupe**

*3. Zotero*

Zotero est un greffon logiciel, ou plug-in, développé par le *Center for History and New Media* de la *George Mason University*. Zotero est à part dans le panorama des logiciels de gestion bibliographique. En effet, ce module s'exécute au sein du navigateur Firefox dont il est une composante optionnelle. Zotero permet de rentrer manuellement des notices bibliographiques au sein d'une base locale, de les classer et de les exporter sous différents formats bibliographiques. Cependant, son réel intérêt est de proposer la possibilité de détecter sur le web des ressources documentaires de différents types, d'en proposer l'enregistrement en masse à la volée en cas de recherche au sein d'une base documentaire ou d'un portail. Dans le cas de références isolées, Zotero permet également de détecter et d'enregistrer un seul élément, comme une page personnelle d'un chercheur qui propose la lecture de son dernier article. La force de ce logiciel est d'autoriser la création et l'export de bibliographies aux formats d'éditeurs de textes les plus repandus avec de plus la mise en forme bibliographique adaptée au domaine de recherche de l'utilisateur, peut être même du type de revue.

*C. Synthèse de bonnes pratiques*

Si nous tenons compte au maximum des leçons offertes par les modèles psychocognitifs de recherche d'informations, il faut respecter quelques règles. Dans l'optique d'offrir à l'usager un maximum de concentration sur sa sélection de documents, nous devons le dégager de la contingence technique. Cela est particulièrement vrai lors de l'intégration d'un document à la bibliographie et de la normalisation du document bibliographique. Lors de l'utilisation de la méthode de Bates, le glanage est particulièrement recommandé. La proximité lexicale entre cueillette et glanage est d'ailleurs frappante. Plutôt que de simplement proposer les métadonnées à l'affiche, nous recommandons donc de les exposer à la détection d'outils pointus tels que le greffon Zotero pour un usage polyvalent. Les excellents Mendeley et JabRef sont à réserver aux usagers de LaTeX. Pour résumer, l'avantage de ces outils est de permettre la détection, l'intégration et l'export de notices pour une production scientifique. Ainsi, trois étapes du processus classique sont facilitées, voire automatisées.

IV. Notre proposition de SRI adapte a l'informatique

Pour donner corps à note étude, jusqu'ici théorique, nous avons créé un SRI spécialisé en informatique scientifique. Nous sommes partis d'une cartographie de la taxonomie informatique ACM CCS[5], que nous avons adaptée en deux langues (français et anglais) à la norme XML graphML[6]. Nous avons établi un cahier des charges strict quant à notre outil qui se devait impérativement de faire progresser les usagers en méthodologie de RI. Notre SRI doit également les aider à maîtriser le domaine de connaissance et les liens internes à la taxonomie (initiation à la sérendipité). Ces choix doivent leur donner confiance en eux en évitant de les placer en situation d'échec comme le bruit ou silence.

---

[5] http://www.acm.org/about/class/1998/
[6] http://graphml.graphdrawing.org/

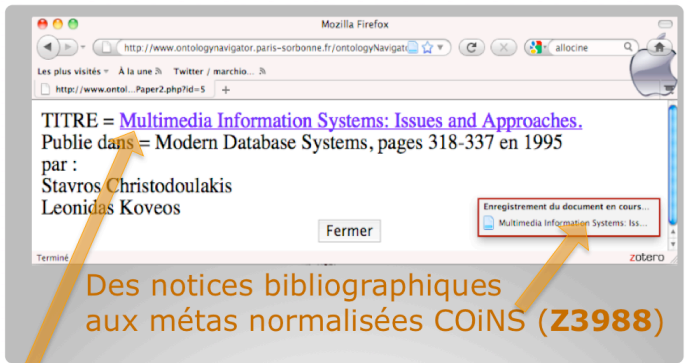

**Figure 3. Enregistrement d'un document dans Zotero**

Nous avons choisi une représentation sous forme de graphe pour mettre en évidence la classification, la hiérarchie des concepts, ainsi que les références croisées. La figure 3 illustre le sens d'utilisation de l'outil et la possibilité de réaliser sa recherche par navigation ou par une requête. Lors de l'usage du moteur de recherche en langue naturelle, français ou anglais, un filtrage du bruit sur l'intitulé est effectué grâce à des listes de mots vides en français et anglais. Une lemmatisation, puis un calcul de proximité statistique de l'ensemble des termes clés de la requête sont réalisés. Cela permet d'identifier une ou plusieurs branches de la taxonomie syntaxiquement proche(s) de la requête. Le positionnement sur un nœud de la taxonomie va dégager des termes clés ou des relations transverses. Le système va ensuite constituer simultanément des requêtes vers la base interne ou les bibliothèques numériques en ligne les plus connues. La base interne est générée et mise à jour depuis les données de référence, issues du travail de Mickael Ley, DBLP[7] [14] avec ses plus de 1,6 million de notices bibliographiques au format BibTeX XML. Dans la base interne, les intitulés des articles correspondants à la requête sont proposés. Un hyperlien vers l'article grâce à Google Scholar est proposé si l' «*url*» est absente de la notice DBLP intégrée. Si l'on utilise les bibliothèques numériques, l'accès aux documents est direct. Les métadonnées sont exposées au format OpenURL / COinS pour une intégration directe au logiciel de gestion bibliographique comme le montre la figure 4. Lorsque l'utilisateur progresse dans sa maîtrise du système, l'abandon du moteur (qui rajoute une étape) deviendra une évidence et la navigation sera systématique.

## CONCLUSION

Dans et article, nous avons établi le désamour des usagers des bibliothèques numériques à vocation scientifique dans l'introduction. La suite a détaillé le processus de recherche d'informations électronique sous les aspects théoriques en sciences psychocognitives, puis pratiques en proposant une étude sur les méthodes, protocoles et outils de recherche d'informations scientifique. Ensuite, nous avons dégagé un ensemble de bonnes pratiques à mettre en œuvre lors de la modélisation d'un SRI. Enfin, nous avons proposé un SRI répondant aux préconisations établies. Dans l'optique dégagée par Bates, nous envisageons de proposer une prévisualisation à facettes de la liste des articles pour optimiser les chances de sérendipité. Il nous reste à confronter *in situ* la validité de ce modèle dans un service commun de documentation universitaire.

---

[7] http://www.informatik.uni-trier.de/~ley/db/